\documentclass[aps,prc,superscriptaddress,twoside,twocolumn,nofootinbib,showpacs]{revtex4}
\usepackage{amsmath,amssymb}
\usepackage{graphicx}
\begin{document}

\title{\boldmath Isobar model analysis of $\pi^0\eta$ photoproduction on protons}
\author{A. Fix}\email{fix@mph.phtd.tpu.ru}
\affiliation{Tomsk Polytechnic University, Tomsk, Russia}
\author{V.L. Kashevarov}\email{kashev@kph.uni-mainz.de}
\affiliation{Institut f\"ur Kernphysik, Johannes Gutenberg-Universit\"at Mainz, Mainz}
\affiliation{Lebedev Physical Institute, Moscow, Russia}
\author{A. Lee}
\affiliation{Tomsk Polytechnic University, Tomsk, Russia}
\author{M. Ostrick}
\affiliation{Institut f\"ur Kernphysik, Johannes Gutenberg-Universit\"at Mainz, Mainz}

\date{today}

\begin{abstract}
Photoproduction of $\pi^0\eta$ on protons in the energy range from threshold to 1.4
GeV is discussed. The data for representative angular distributions recently obtained at
MAMI~C are analyzed using an isobar model. The isobars considered are $\Delta(1232)$ and
$S_{11}(1535)$ for $\pi^0 p$ and $\eta p$ states, respectively. In accordance with the
results of earlier works the main features of the reaction are explained through the
dominance of the $D_{33}$ wave with a relatively small admixture of positive parity
resonances. Comparison with recent experimental results for the photon beam asymmetry is
carried out.
\end{abstract}

\pacs{25.20.Lj, 
      13.60.Le, 
      14.20.Gk  
      } %

\maketitle



%
\section{Introduction}
Several important features of $\pi^0\eta$ photoproduction on protons were explained in
earlier papers \cite{Dor,Dor2,Horn,Ajaka}, where different versions of an isobar model
were used as a theoretical basis. According to the experimental results of
\cite{Horn,Ajaka,Tohoku} the cross section for $\gamma p\to\pi^0\eta p$ exhibits quite a
rapid rise at low energies and reaches its characteristic value $\sigma\approx 3.5$\,$\mu
b$ at $E_\gamma=1.4$\,GeV, about 0.5 GeV above threshold. Such a behavior might indicate
that the cross section is mainly governed by the increasing phase space so that it is
dominated by $s$-waves. As has been noted in \cite{Dor,FOT} the Born terms seem to be
insignificant and the reaction is expected to proceed via formation of baryon resonances
decaying into $\pi\eta N$.

Within an isobar model, production of two mesons may conveniently be described in terms
of intermediate quasi-two-body states, the isobars. These states naturally appear if the
attractive two-body interaction between the final particles lead to narrow resonances. In
the case of the $\pi\eta N$ state the $\pi N$ and $\eta N$ isobars are identified with the
$\Delta(1232)$ and $S_{11}(1535)$ which are well known to dominate $\pi N$ and $\eta N$
scattering. In such a picture the observed growth of the total cross section would point
to the existence of baryon resonances, belonging to the third resonance region, which may
decay into $s$-wave $\eta\Delta$ or $\pi S_{11}$ configurations. One can easily show that
only in the $D_{33}$ ($P_{31}$) state the $\eta\Delta$ ($\pi S_{11}$) system can be
produced in the relative $s$-wave.

The model for $\gamma N\to\pi^0\eta N$, with the resonance \\ $D_{33}(1700)$
dominating the amplitude, has been developed in ref.\,\cite{Dor}, where $\pi^0\eta$ and
$\pi^0 K^0$ photoproduction were studied within a coupled channel approach. The authors
used the $D_{33}\to\eta\Delta$ decay, treated on a tree level, as a main driving
mechanism. The crucial importance of $D_{33}$ in this calculation is ultimately a
consequence of quite a large (complex) coupling constant $g_{\eta\Delta}$ whose value
$g_{\eta\Delta}=1.7-1.4i$ was taken from \cite{Sarkar}. The production of the $\pi
S_{11}$ state is calculated microscopically as a series of interactions
$\eta\Delta\to(\pi\eta N,\,\pi K\Lambda,\,\pi K\Sigma)\to\pi S_{11}$ taken up to the
first order in the two-body $t$-matrices for $\eta N$ and $K\Lambda\,(K\Sigma)$
scattering.

On the experimental side, recent studies \cite{Horn,Ajaka,MAMI} have also implied that
the resonance $D_{33}(1700)$ plays a crucial role in $\pi^0\eta$ photoproduction. In
ref.\,\cite{Horn} the data for $\gamma p\to\pi^0\eta p$ were explained using the PWA
method \cite{Saran} which gave definite quantitative conclusions about contributions of
different $N$- and $\Delta$-like resonances. In particular, the authors reported that at
higher energies ($W>1.8$\,GeV) the $D_{33}$ wave remains important and is presumably
populated by the $D_{33}(1940)$ resonance.

In ref.\,\cite{Ajaka}, in addition to the basic cross section, the linear beam asymmetry
$\Sigma$ was measured. In common with the results of \cite{Horn} the authors of
\cite{Ajaka} emphasized the dominance of the $D_{33}(1700)$. The importance of the
$\eta\Delta$ configuration is indicated by the pronounced peak in the $\pi^0p$ mass
spectra around the $\Delta$ mass. The data for the $\Sigma$-asymmetry are in general
agreement with the predictions of the model~\cite{Dor}. The $D_{33}(1700)$ dominance is
deduced from the fact that without this resonance one obtains strong qualitative
disagreement with the measured asymmetry.

A remarkable feature of the data reported in ref.\,\cite{Ajaka} is that the $\pi S_{11}$
configuration appears to not be as pronounced as it may follow from the theoretical
calculation of ref.\,\cite{Dor}. The case in point is that the apparent shifting of the
$\pi^0 p$ peak predicted by the model \cite{Dor} is not evident from the experimental
results. Instead of this, as is shown in \cite{Ajaka}, the simple model containing only
$\eta\Delta$ decay accounts for the data much better than the sophisticated approach
including $\pi S_{11}$ formation. On the other hand, trivial assumption that the role of
$\pi S_{11}$ configuration in $\gamma p\to\eta\pi^0 p$ is negligible may lead to
misinterpretation of the data. It is clear that the $S_{11}(1535)$ resonance should, in
any case, be present on the Dalitz plot due to the $\eta N$ final state interaction. The
reason for its 'disappearance' may lie in the fact that the $\eta N$ scattering cross
section does not exhibit a true resonance behavior due to the closeness of the resonance
mass to the $\eta N$ threshold. Furthermore, the presence of an intensive $D_{33}\to\eta
\Delta$ decay makes it difficult to estimate quantitatively the role of the $\pi S_{11}$
channel. These observations suggest that studying $\pi N$ or $\eta N$ invariant mass
distributions does not enable us to clarify the separation of the events related to $\pi
S_{11}$ formation. In this connection, it is desirable to have methods allowing
identification of $\pi S_{11}$ without recourse to the $\eta N$ invariant mass
distribution.

The crucial role of the $D_{33}$ configuration in $\pi^0\eta$ photoproduction is a very
important result by itself. It implies that, apart from the $(\gamma,\pi^0)$ and
$(\gamma,\eta)$ reactions determined respectively by the $\Delta$ and $S_{11}$ formation,
we have another process, $(\gamma,\pi^0\eta)$, whose amplitude in a wide energy range is
mainly governed by a single dynamical mechanism, the excitation of the $D_{33}$ wave, in
particular the $D_{33}(1700)$ state. Clearly, this feature provides an effective method
to study the properties of this resonance. However, in contrast to single $\pi^0$ or
$\eta$ photoproduction, here we face the technical problems associated with three-body
kinematics, where the particle energies and angles are distributed continuously. It
therefore becomes difficult to perform a general multipole analysis, primarily since
there is a multitude of ways to couple angular momenta of particles to the total angular
momentum.

Important steps on the road to a systematic study of the partial wave structure of
$\pi^0\eta$
photoproduction were made in ref.\,\cite{FOT}, where the authors presented
a phenomenological model for $\gamma N\to\pi^0\eta N$ and discussed the angular distributions
of final particles. Although, in general, unpolarized measurements are not
sufficient to distinguish between all partial waves, assuming that only one of these
waves is important, it is possible to reach a rather definite conclusion about the
corresponding quantum numbers. The analysis permits the identification of the quantum
numbers of the dominant resonance states simply by ana\-ly\-sing the shape of the
measured angular dependencies, in which the states with different spin-parities exhibit
their own signatures. It is also emphasized in \cite{FOT} that some of the observables
depend weakly on the model parameters. This fact makes such a method especially useful.

The formalism developed in ref.\,\cite{FOT} was then applied to the analyses of the data
recently obtained at MAMI~C~\cite{MAMI}. As a main subject, the authors of \cite{MAMI}
studied the angular distributions of particles in $\gamma p\to\pi^0\eta p$. This approach was
shown to be very effective as a tool to study the properties of the $D_{33}(1700)$
resonance. In particular, the ratio of $\pi S_{11}$ to $\eta\Delta$ decay widths, which,
as pointed out above, is difficult to extract from the invariant mass distribution, was
deduced from the data with appreciated accuracy.

Here we continue the analysis of the reaction $\gamma N\to\pi^0\eta N$ using the isobar model
which was partially described in refs.\,\cite{FOT} and \cite{MAMI}. The paper is
organized as follows. In the next section we briefly outline the formalism, and summarize
some results, of ref.\,\cite{FOT} which provides a framework for the calculation to follow.
Then in Sect.~\ref{results} we present our results where the main emphasis is put on the
angular distributions of the final particles. In Sect.\,\ref{BeamAsmtry} we consider in
some detail the photon asymmetry $\Sigma$ for $\gamma p\to\pi^0\eta p$ and discuss the main
properties of this observable. In Sec.\,\ref{Conclusion} we close with a brief summary
and an outlook.


\section{Theoretical basis}\label{basis}

We consider the reaction
\begin{eqnarray}\label{10}
&&\gamma(\omega_\gamma,\vec{k}\,)+N_i(E_i,-\vec{k})\\
&&\phantom{xxxxxxxx}\to\pi^0(\omega_\pi,\vec{q}_\pi)+\eta\,(\omega_\eta,\vec{q}_\eta)+N_f(E_f,\vec{p}_f)\,\nonumber
\end{eqnarray}
in the overall center-of-mass system. The corresponding energies and three-momenta of the
particles are given in the parentheses. Furthermore, throughout the paper we use the
following notations for kinematic variables:
\begin{tabbing}
x \= xxxxxxxxxxxxxxxxxxxxxxx \= xxxxxxxxxxxxxxxxxxxxxxxxxxxxxxxxxxxxxxxxxxxxxxxxxxxxxxxxx
\kill
\> $E_\gamma$ \> photon lab energy,  \\
\> \> \\
\> $W$ \> total c.m.\ energy,  \\
\> \> \\
\> $\omega_{\alpha}$,\ \ \ $\alpha=\pi N,\,\eta N$ \> invariant energy of the \\
\> \> two-particle subsystem $\alpha$, \\
\> \> \\
\> $\Omega_i=(\Theta_i,\Phi_i)$,\ \ \ $i=\pi,\eta$ \> solid angle of the momentum
\\
\> \> $\vec{q}_i$ in the overall c.m.\ frame,\\
\> \> \\
\> $\omega^*_i,\ \vec{q}^{\,*}_i$,\ \ \ $i=\pi,\eta$ \> energy and three-momentum\\
\> \> of the $i$-th particle\\
\> \> in the $iN$ c.m.\ frame,\\
\> \> \\
\> $\Omega^{K(H)}_i=(\theta^{K(H)}_i,\phi^{K(H)}_i)$, \> solid angle of the \\
\> $i=\pi,\eta$ \> momentum $\vec{q}^{\,*}_i$, related \\
\> \> to the $K$($H$) system.
\end{tabbing}
The meaning of $K$ and $H$ systems is explained by Fig.\,\ref{fig2} (see also text after
eq.\,(\ref{30a})).

\begin{figure}
\begin{center}
\resizebox{0.47\textwidth}{!}{%
\includegraphics{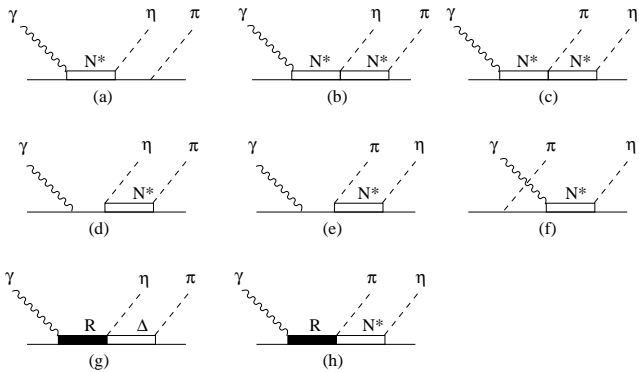}}
\caption{Diagrams representing the amplitude for the $\gamma N\to\pi\eta N$ reaction in a
simple isobar model. The notations $\Delta$ and $N^*$ are used for the resonances
$\Delta(1232)$ and $S_{11}(1535)$.} \label{fig1}
\end{center}
\end{figure}

\begin{figure}
\begin{center}
\resizebox{0.45\textwidth}{!}{%
\includegraphics{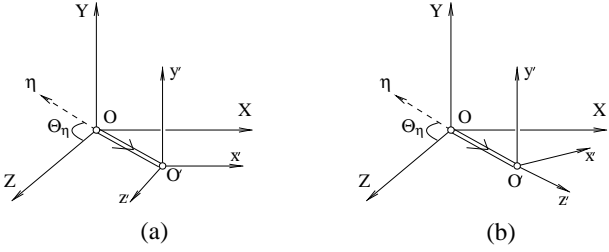}}
\caption{The coordinate systems $(x'y'z')$ used for the analysis of angular distributions
of pions in the $\pi p$ rest frame. In the canonical system ($K$ system) the $z'$ axis is
taken parallel to the beam direction, whereas in the helicity system ($H$ system) it is
aligned along the total $\pi p$ momentum. In both systems, the $x'$ axis is in the
reaction plane and the $y'$ axis is chosen as
$\hat{y}'=(\vec{p}_\eta\times\vec{k}_\gamma)/|\vec{p}_\eta\times\vec{k}_\gamma|$.}
\label{fig2}
\end{center}
\end{figure}

A theoretical basis of our calculation is presented in ref.\,\cite{FOT} and partially in
ref.\,\cite{MAMI}. We assume that the $\pi\eta$ system does not resonate over the energy
region in question. The reaction amplitude is divided into the Born and the resonance
part,
\begin{equation}
t_{m_f\lambda}=t_{m_f\lambda}^B+t_{m_f\lambda}^R,
\end{equation}
and is visualized by means of the diagrams in Fig.\,\ref{fig1}. The indices $m_f=\pm 1/2$
and $\lambda=\pm 1/2$, $\pm 3/2$ denote the $z$-projection of the final
nucleon spin and the initial state helicity, respectively. The Born sector is formed by
the diagrams (a)-(f). As already noted, its role in the channel $\pi^0\eta p$ turns out
to be insignificant.

According to the isobar model concept, the resonance part of the amplitude may be
superposed in the usual fashion, as a sum of two terms \footnote{In the expressions to
follow the resonance $S_{11}$(1535) is denoted by $N^*$.}
\begin{equation}\label{15}
t^R_{m_f\lambda}=t^{(\eta\Delta)}_{m_f\lambda}+t^{(\pi N^*)}_{m_f\lambda}\,
\end{equation}
corresponding to the diagrams (g) and (h) in Fig.\,\ref{fig1}. Each term in
eq.\,(\ref{15}) has the form
\begin{eqnarray}\label{20}
&&t_{m_f\lambda}^{(\alpha)}(W,\vec{q}_\pi,\vec{q}_\eta,\vec{p}_f)=\sum_{R(J^\pi;T)}C_T
A^R_\lambda(W)\,G_R(W)\nonumber\\
&& \phantom{xxx}\times f^{R(\alpha)}_{m_f\lambda}
(W,\vec{q}_\pi,\vec{q}_\eta,\vec{p}_f)\,,\quad \alpha=\eta\Delta,\,\pi N^*\,,
\end{eqnarray}
where the summation is over the resonance states determined by their spin-parity $J^\pi$
and isospin $T$. The latter is incorporated through the factor $C_T$. Since the $\eta$-meson
has isospin zero, this factor is the same for both the $\eta\Delta$ and $\pi N^*$
channels. In our case, where only the states with $T=3/2$ are assumed to contribute,
$C_T$ is 2/3.

The functions $A_\lambda^R(W)$ are the helicity amplitudes for $\gamma p\to R$. Their
energy dependence was parametrized in the simple form
\begin{equation}\label{AW}
A^R_\lambda(W)=A^R_\lambda(M_R)\left(
\frac{\omega_\gamma}{\omega_\gamma^R}\right)^{n_\lambda}\,,
\end{equation}
where $M_R$ is the resonance mass and $\omega_\gamma^R$ is the c.m.\ photon energy,
calculated at the resonance position. For the dominant resonance, $D_{33}(1700)$, the
exponent $n_{1/2}$ was set equal to 1, whereas the function $A^{D_{33}}_{3/2}(W)$ was
fitted to the data separately in each energy bin in Fig.\,\ref{fig4}. For the other
resonances we used $n_\lambda=0$.


The factors $G_R(W)$ stand for the resonance propagators for which we use the
nonrelativistic Breit-Wigner form
\begin{equation}\label{25}
G_R(W)=\frac{1}{W-M_R+\frac{i}{2}\Gamma^R_{tot}(W)}\,.
\end{equation}
In eq.\,(\ref{20}), the functions $f^{R(\alpha)}_{m_f\lambda}$, $\alpha=\eta\Delta,\,\pi
N^*$, describe the decay of the resonance $R$ into $\pi\eta N$ via intermediate formation of
$\eta\Delta$ and $\pi N^*$ states.

We take the $Z$-axis of the overall c.m.\
frame along the photon momentum $\vec{k}$. The $OXZ$ plane is spanned by the momenta
$\vec{k}$ and $\vec{q}_\eta$ (see Fig.\,\ref{fig2}). To fix the final state kinematics we
choose as independent variables the particle momenta associated with the partition $\eta
+(\pi N)$, the solid angle of the $\eta$ momentum
$\Omega_\eta=(\Theta_\eta,\Phi_\eta=0)$, the invariant energy $\omega_{\pi N}$ of $\pi N$
subsystem, and the solid angle of the momentum $\vec{q}_\pi^{\,*}$ in the $\pi N$ rest
frame.

For the data analysis, two types of coordinate systems $O'x'y'z'$, related to the $\pi N$
c.m.\ frame, are used. In the first one, the canonical ($K$) system
(Fig.\,\ref{fig2}(a)), the axes are chosen parallel to those in the main c.m.\ frame
$OXYZ$ fixed in space. In the helicity ($H$) system (Fig.\,\ref{fig2}(b)) the $z'$-axis
is taken to be along $\vec{q}_\pi+\vec{p}_f=-\vec{q}_\eta$. For a single event the
systems are connected via appropriate rotation by the angle $\pi-\Theta_\eta$ around the
$y'$-axis. After integrating over $\Theta_\eta$ is carried out this equivalence does not
hold, and in this sense the distributions $W(\Omega^K_i)$ and $W(\Omega^H_i)$,
$i=\pi,\eta$ are independent. The values related to the $K$ and $H$ frames are further
denoted by the indices $K$ and $H$ respectively.

In the $K$ system, the pion momentum, $\vec{q}_\pi^{\,*}$, is related to the pion
momentum $\vec{q}_\pi$ in the overall c.m.\ frame via the Lorentz boost along the vector
$-\vec{q}_\eta$, giving
\begin{equation}\label{30}
\vec{q}_\pi=\vec{q}_\pi^{\,*}+X_\pi\vec{q}_\eta\,,\quad
X_\pi=-\frac{\omega_\pi^*+\omega_\pi}{\omega_{\pi N}+W-\omega_\eta}\,.
\end{equation}

In Sect.\,\ref{BeamAsmtry} we also use the partition $\pi+(\eta N)$, taking as independent
kinematical variables the solid angle $\Omega_\pi=(\Theta_\pi,\Phi_\pi=0)$, the invariant
energy $\omega_{\eta N}$, and the spherical angles of the $\eta$ momentum in the $\eta N$
c.m.\ frame. In this case the reaction plane is spanned by the vectors $\vec{k}$ and
$\vec{q}_\pi$. The relation between the momentum of the $\eta$ meson is similar to that in
(\ref{30}):
\begin{equation}\label{30a}
\vec{q}_\eta=\vec{q}_\eta^{\,*}+X_\eta\vec{q}_\pi\,,\quad
X_\eta=-\frac{\omega_\eta^*+\omega_\eta}{\omega_{\eta N}+W-\omega_\pi}\,.
\end{equation}

We adhere to the non-relativistic concept of angular momentum and do not include, for
instance, effects of spin transformation caused by the Lorentz boost. Then, in the
coordinate system chosen, the angular dependence of the amplitudes (\ref{20}) may be
decomposed by means of spherical functions. In the $K$ system, associated with the
partition $\eta+(\pi N)$ (Fig.\,\ref{fig2}) we will have
\begin{eqnarray}\label{35a}
&&f^{R(\eta\Delta)K}_{m_f\lambda}= F^{R(\eta\Delta)}(\omega_{\pi N})\sum_{m}C^{\frac32M_\Delta}_{1m\,\frac12m_f}\nonumber\\
&&\phantom{xxxx}\times
C^{J\lambda}_{L_\eta M_\eta\,\frac32M_\Delta}\,Y_{1m}(\Omega^K_\pi)\,d^{L_\eta}_{M_\eta 0}(\Theta_\eta)\,,\\
\label{35b} &&f^{R(\pi N^*)K}_{m_f\lambda}= F^{R(\pi N^*)}(\omega_{\eta N})\
C^{J\lambda}_{L_\pi
M_\pi\frac12m_f}\sum\limits_{l=0}^{L_\pi}A_l\nonumber\\
&&\phantom{x}\times\sum_m C_{L_\pi-l\,M_\pi-m\ lm}^{L_\pi
M_\pi}Y_{lm}(\Omega_\pi^K)\,d^{L_\pi-l}_{M_\pi-m\,0}(\Theta_\eta)\,.\ \quad
\end{eqnarray}
Here, $C_{j_1m_1\ j_2m_2}^{j_3m_3}$ are the usual Clebsch-Gordan coefficients for the
coupling $\vec{j}_1+\vec{j}_2=\vec{j}_3$. The coefficients $A_l$, determined as
\begin{equation}
A_l=\left(\frac{X_\pi
q_\eta}{q^*_\pi}\right)^l\sqrt{\frac{(2L_\pi-1)\,(2L_\pi)!}{(2l-1)\,(2L_\pi-2l)!\,(2l)!}}\,,
\end{equation}
stem from the expansion of the function $Y_{L_\pi M_\pi}(\Omega_\pi)$ over the products
of the spherical functions depending on $\Omega_\pi^K$ and $\Omega_\eta$.

The corresponding expressions for $f^{R(\alpha)K}_{m_f\lambda}$ can easily be obtained
from eqs.\,(\ref{35a}) and (\ref{35b}) via rotation of $Y_{lm}(\Omega_\pi^K)$ by the
angle $\Theta_\eta$ around the $Y$ axis.

The coefficients $F^{R(\alpha)}(\omega)$, $\alpha=\eta\Delta,\,\pi N^*$ are parametrized
in terms of coupling constants
\begin{eqnarray}\label{40a}
&&F^{R(\eta\Delta)}(\omega_{\pi N})=\frac{f_{R\eta\Delta}f_{\Delta\pi
N}}{m_\pi^{L_\eta+1}}\ G_\Delta(\omega_{\pi N})\,q^{L_\eta}_\eta\, q_\pi^*,\\
&& \nonumber\\
\label{40b} &&F^{R(\pi N^*)}(\omega_{\eta N})=\frac{f_{R\pi N^*}f_{N^*\eta
N}}{m_\pi^{L_\pi}}\ G_{N^*}(\omega_{\eta N})\,q^{*\,L_\pi}_\pi,\ \quad\
\end{eqnarray}
where $G_\Delta$ and $G_{N^*}$ are the $\Delta$ and $N^*$ isobar propagators.

Unlike ref.\,\cite{Dor} here we use a phenomenological rather than a dynamical approach
to describe the transition to $\pi N^*$. Namely, the vertex $R\to\pi N^*$ appearing in
the diagram (h) in Fig.\,\ref{fig1} was parametrized in the form of an ordinary resonance
decay $R\to\pi N^*$ fixed by the constant $f_{R\pi N^*}$ (see eq.~(\ref{40b})). In
contrast, in ref.\,\cite{Dor} only the $\eta\Delta$ decay is treated as a 'primary
process' whereas $\pi N^*$ is generated during the $\eta p$ final state interaction. As
noted in \cite{FOT}, we expect our approximation to be reasonable at least around the
resonance maximum.

\begin{table}
\renewcommand{\arraystretch}{2.0}
\caption{Angular momenta associated with a decay of the resonance $R(J^\pi)$  into the
channels $\eta\Delta$ (denoted by $L_\eta$) and $\pi N^*$ ($L_\pi$). } \label{ta1}
\begin{center}
\begin{tabular*}{8cm}
{@{\hspace{0.6cm}}c@{\hspace{0.6cm}}|@{\hspace{1.6cm}}c @{\hspace{1.6cm}}c}
\hline\noalign{\smallskip}
$J^\pi(L_{2T2J})$ & $L_\eta$ & $L_\pi$ \\
\noalign{\smallskip}\hline\noalign{\smallskip}
$\frac12^-(S_{31})$  & 2  & 1 \\
$\frac12^+(P_{31})$  & 1  & 0 \\
$\frac32^-(D_{33})$  & 0,\ 2 & 1 \\
$\frac32^+(P_{33})$  & 1,\ 3 & 2 \\
$\frac52^-(D_{35})$  & 2,\ 4 & 3 \\
$\frac52^+(F_{35})$  & 1,\ 3 & 2 \\
\noalign{\smallskip}\hline
\end{tabular*}
\end{center}
\end{table}

The angular distributions are analysed in terms of the functions $W(\theta)$ and $W(\phi)$
which are differential cross sections normalized to unity. For instance,
\begin{eqnarray}\label{55}
&&W(\theta_\pi^K)=\frac{1}{\sigma}\int\limits_0^{2\pi}\frac{d\sigma}{d\Omega_\pi^K}\,d\phi_\pi^K\,,\\
\label{55a}
&&W(\phi_\pi^K)=\frac{1}{\sigma}\int\limits_0^\pi\frac{d\sigma}{d\Omega_\pi^K}\,\sin\theta_\pi^K
d\theta_\pi^K\,,
\end{eqnarray}
with $\sigma$ standing for the total cross section.

To effectively include the total cross section into the fitting procedure we use the
products $\bar{\sigma}W(\theta^{K(H)}_\pi)$ and $\bar{\sigma}W(\phi^{K(H)}_\pi)$, where
$\bar{\sigma}$ is the total cross section, averaged over the corresponding energy bin.

The adjustable parameters are collected in Table\,\ref{ta2}. Since the resonance
amplitudes are proportional to the product of the electromagnetic and hadronic couplings,
the amplitudes $A_\lambda$ and the ratios $\beta_\alpha=\Gamma_{\pi\eta
N}^{(\alpha)}/\Gamma^R_{tot}$, $\alpha=\eta\Delta,\,\pi N^*$, cannot be well determined
individually. For this reason, we use the quantity $\sqrt{\beta_{\eta \Delta}}A_{1/2}$
together with the ratio of the helicity amplitudes $A_{3/2}$ and $A_{1/2}$, and the ratio
$r$ of $R\to\pi N^*$ and $R\to\eta\Delta$ branchings.


We note that our fitting procedure did not include varying the total widths. The reason
lies in the closeness of the resonances, especially $D_{33}(1700)$, to the $\pi\eta$
production threshold. As a result, the set of angular distributions alone does not impose
sufficiently stringent constraints on the values of the resonance widths, and to some
extent, on their masses. In this respect, we fix the widths, taking their values from the
PDG compilation~\cite{PDG} or from the references cited there (the width of
$D_{33}(1700)$ was taken from the recent analysis of~\cite{Arndt06} and that of
$P_{31}(1750)$ from ref.\,\cite{Manley}). For the same reason, the masses of the well
known resonances, rated by four or three stars, were varied around their values given by
PDG\,\cite{PDG}.


The orbital angular momenta $L_\eta$ and $L_\pi$ related to the $\eta\Delta$ and $\pi
N^*$ decays of the resonances are listed in Table~\ref{ta1}. As discussed in
Sect.\,\ref{Angular distributions}, in the $D_{33}(1700)\to\eta\Delta$ decay we included
both $s$- and $d$-waves. As for other resonance states, for the sake of simplicity, we
take only the lowest of the two possible values of $L_\eta$ (see Table\,\ref{ta1}).

\begin{figure}
\begin{center}
\resizebox{0.46\textwidth}{!}{%
\includegraphics{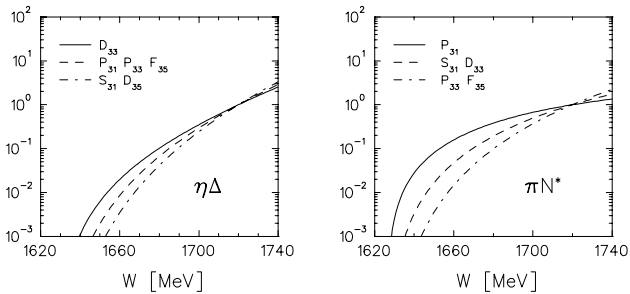}}
\caption{Powers of $\eta\Delta$ and $\pi N^*$ relative momenta averaged over the isobar
distribution as functions of the total c.m.\ energy. As indicated in the legend, the
curves determine energy dependence of partial decay widths of the corresponding
resonances (see eqs.\,(\ref{65a}) and (\ref{65b})). As noted in text, for
$R\to\eta\Delta$ decay only the lowest value of the orbital momenta $L_\eta$ is
considered.} \label{fig3}
\end{center}
\end{figure}

\section{Discussion of the results}\label{results}

The general properties of the measured angular distributions and energy spectra were
considered in ref.\,\cite{MAMI}. To interpret the data, the authors of ref.\,\cite{MAMI}
used the single resonance model. It was assumed that there is one resonance (or, in
general case, one partial wave) dominating the amplitude in a wide energy region.
Following this concept they retain only one term in the sum (\ref{20}) to roughly
reproduce the main features of the observed angular distributions and thus to reach a
tentative conclusion about the quantum numbers of this dominant resonance.

The results obtained in ref.\,\cite{MAMI} are now used as a starting point for a more
refined analysis. Namely, having established the appropriate magnitude of the dominant
partial wave, we successively add other waves in order to reproduce the structural
details of the measured quantities. Thereupon we calculate the angular distributions
measured in ref.\,\cite{Horn}, as well as the photon beam asymmetry presented in
\cite{Ajaka} and \cite{Gutz}, and compare our results with the data.


\subsection{Angular distributions}\label{Angular distributions}

Before dealing with the results of our fit, it is instructive to qualitatively discuss
the role of different partial waves in the angular distributions obtained in the
experiment. Firstly, we consider the near-threshold region where the energy is not high
enough to produce isobars in higher angular momentum states, so that spin-parity
selection rules may be used at least for qualitative estimations. In essence, the
question is related to the energy dependence of the partial decay widths for
$R\to\alpha\to\pi\eta N$, $\alpha=\eta\Delta,\,\pi N^*$, which in turn are determined by
the appropriate powers of the $\eta\Delta$ and $\pi N^*$ relative momenta averaged over
the isobar distribution in the available phase space:
\begin{eqnarray}\label{65a}
&&\langle q_\eta^{2L_\eta+1}\rangle(W)=\frac{1}{2\pi}\int
\limits_{M_N+m_\pi}^{W-m_\eta}q_\eta^{2L_\eta+1}(W,\omega)\nonumber\\
&&\phantom{xxxxx}\times\frac{
\Gamma_{\Delta}(\omega)}{(\omega-M_\Delta)^2+\Gamma_\Delta(\omega)^2/4}\ \omega\,d\omega\,,\\
\label{65b} &&\langle q_\pi^{2L_\pi+1}\rangle
(W)=\frac{1}{2\pi}\int\limits_{M_N+m_\eta}^{W-m_\pi}q_\pi^{2L_\pi+1}(W,\omega)\nonumber\\
&&\phantom{xxxxxx}\times \frac{
\Gamma_{N^*}(\omega)}{(\omega-M_{N^*})^2+\Gamma_{N^*}(\omega)^2/4}\ \omega\,d\omega\,.
\end{eqnarray}

The particle momenta appearing in eq.\,(\ref{65a}) and (\ref{65b}) are determined by the
usual kinematic formula
\begin{equation}\label{50}
q_i(W,\omega)=\frac{\lambda^{1/2}(W,\omega,m_i)}{2W}\,,\quad i=\pi,\eta
\end{equation}
with $\lambda$ standing for the triangle function $\lambda(x,y,z)=(x-y-z)^2-4yz$.

In Fig.\,\ref{fig3} we present the ratio $\langle q_i^{2L_i+1}\rangle(W)/\langle
q_i^{2L_i+1}\rangle(M)$, $i=\pi,\eta$, calculated for $M=1.72$\,GeV with typical isobar
masses and widths:
\begin{eqnarray}\label{70}
&&M_\Delta=1232\,\mbox{MeV}\,,\quad \Gamma_\Delta(M_\Delta)=120\,\mbox{MeV}\,,\nonumber\\
&& \nonumber\\
&&M_{N^*}=1535\,\mbox{MeV}\,,\quad \Gamma_{N^*}(M_{N^*})=150\,\mbox{MeV}\,,\nonumber\\
&& \nonumber\\
&&\Gamma_{N^*\eta N}(M_{N^*})=45\%\,\Gamma_{N^*}(M_{N^*})\,.\nonumber
\end{eqnarray}
As might be expected, in each individual channel, $\eta\Delta$ or $\pi N^*$, the energy
dependence of the partial cross sections exhibits the typical threshold behavior
$\sigma_{L_i}\sim \langle q^{2L_i+1}\rangle$, $i=\pi,\,\eta$. This simple relation,
however, is slightly violated in the real situation when both configurations,
$\eta\Delta$ and $\pi N^*$, are included. Firstly, because of the different parities of
$\Delta$ and $N^*$ the values $L_\eta$ and $L_\pi$, entering the same partial wave
$J^\pi$, always differ by 1. Secondly, the nominal thresholds corresponding to the
production of stable $\Delta$ and $N^*$ are $E_\gamma=1.22$ and $1.01$ GeV, respectively,
for $\eta\Delta$ and $\pi N^*$ channels. As a result, the $\eta\Delta$ channel is
effectively opened at energies about 200 MeV higher than those associated with $\pi N^*$
production. Furthermore, the smaller mass of the pion, in comparison to the $\eta$, leads
to the availability of a larger phase space for the $\pi N^*$ configuration already at
rather low energies. By referring to Fig.\,\ref{fig3} it may be seen that in the
$\eta\Delta$ channel the $D_{33}$ state exhibits the largest rate of growth. The next
most important waves are $P_{31}$ and $P_{33}$, producing the $\eta\Delta$ system in the
relative $L_\eta=1$ state. Furthermore, the $P_{31}$, leading to the $s$-wave in the $\pi
N^*$ mode, may kinematically be the most favorable wave close to the threshold.

\begin{figure}
\begin{center}
\resizebox{0.32\textwidth}{!}{%
\includegraphics{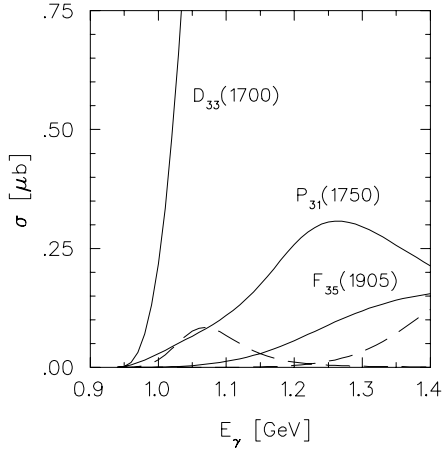}}
\caption{Estimate of resonance contributions to $\gamma p\to\pi^0\eta p$ according to
formula (\ref{15}). Presented are the resonances which were examined in the quark model
analysis of ref.\,\cite{Roberts} as the baryons decaying into the $\eta\Delta$ channel.
The dashed lines show the contribution of $P_{33}(1600)$ and $P_{33}(1920)$. For the
electromagnetic and hadronic decay widths the values from the PDG compilation~\cite{PDG}
are taken. For the partial widths $\Gamma_{\pi\eta N}$ the same relation $\Gamma_{\pi\eta
N}=0.1\,\Gamma_{tot}$ for each resonance is used. } \label{fig4a}
\end{center}
\end{figure}


Some $\Delta$ resonances, which may be coupled with the $\eta\Delta$ state, were
calculated within the constituent quark model of ref.\,\cite{Roberts}. A rough estimation
of their contribution can be made by taking the corresponding parameters from the
Particle Data Tables~\cite{PDG}. Neglecting background and using unitarity one arrives at
the sum over the resonance states $R$
\begin{equation}\label{250}
\sigma(W)\approx\sum\limits_R \sigma^R(W)\,,
\end{equation}
with the partial cross section reading
\begin{equation}\label{250a}
\sigma^R(W)=\frac{\pi}{\omega_\gamma^2}(2J_R+1)\frac{\Gamma^R_{\pi\eta
N}(W)\Gamma^R_{\gamma N}(W)}{(W-M_R)^2+\Gamma^{R\,2}_{tot}(W)/4}\,.
\end{equation}
The cross section predicted by formula (\ref{250}) is plotted in Fig.\,\ref{fig4a}
together with the contributions of the individual resonances. For the partial decay widths we
used the common relation $\Gamma_{\pi\eta N}(M)$ $=0.1\,\Gamma_{tot}(M)$ for all states.
Although this estimate is very rough because of the uncertainties in the partial widths
$\Gamma_{\pi\eta N}$, one can expect that the $P_{33}(1600)$ resonance
already mentioned can really be visible near the threshold, whereas the $P_{31}(1750)$ and
$F_{35}(1905)$ may
come into play at higher energies.

\begin{figure*}
\begin{center}
\resizebox{1.0\textwidth}{!}{%
\includegraphics{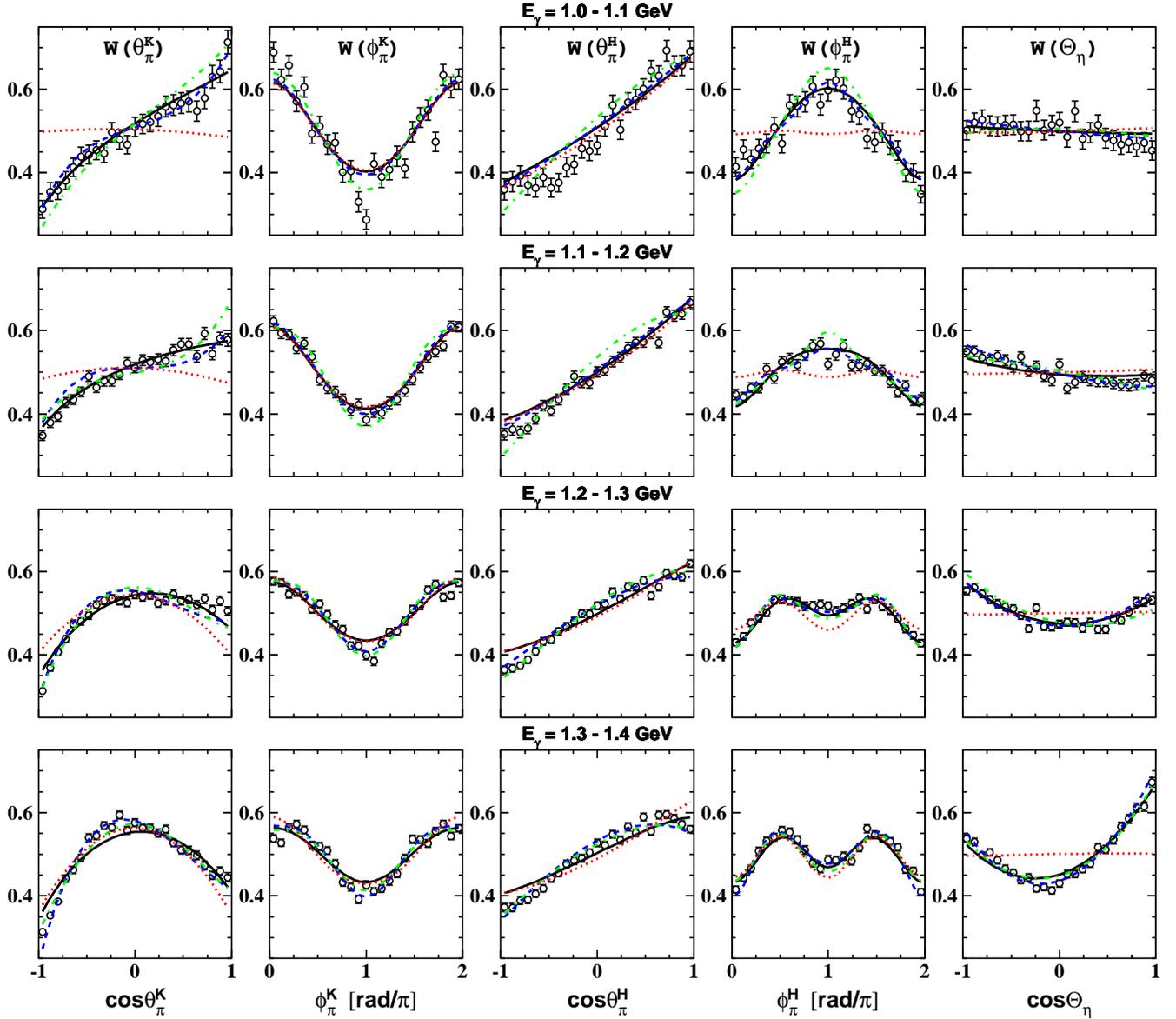}}
\caption{Angular distributions of pions calculated in the canonical and helicity
coordinate systems (the first four columns) and angular distribution of $\eta$ mesons in
the overall $\gamma p$ c.m. frame (the fifth column). The data are taken from
\cite{MAMI}. The curves are our isobar model calculation corresponding to the solutions
presented in Table\,\ref{ta2}: solution I (solid), solution II (dashed), solution
III (dash-dotted). The dotted curve is obtained within the single $D_{33}$ model, including
both $D_{33}(1700)$ and $D_{33}(1940)$ resonances. In the last case the parameters of
both resonances were taken from the solution I.}
\label{fig4}
\end{center}
\end{figure*}

Since we expect the $D_{33}$ to dominate, it is useful to have an analytic expression for
the cross section containing only $D_{33}$. Retaining in the sum (\ref{20}) the term with
$J^\pi=3/2^-$, and using in eqs.\ (\ref{35a}) and $(\ref{35b})$  $L_\eta=0$ (for the
moment we omit the $d$-wave in the $D_{33}\to\eta\Delta$ decay) and $L_\pi=1$ from
Table~\ref{ta1}, we obtain in the $H$ system
\begin{eqnarray}\label{75}
&&\frac{d\sigma}{d\omega_{\pi N}d\Omega^H_\pi d\cos\Theta_\eta}\sim \nonumber\\
&&\phantom{xx}\frac{2}{3}\left(A_{1/2}^2+A_{3/2}^2\right)
\Big(|c_1|^2+|c_2|^2+2Re\,(\bar{c}_1c_2)\cos\theta_\pi^H\Big)
\nonumber\\
&&\phantom{x}+\,\frac{1}{2}\left(A_{1/2}^2-A_{3/2}^2\right)\Big[\,
\frac{2}{3}\left(1-3\cos^2\Theta_\eta\right)\nonumber\\
&&\phantom{x}\times \Big(|c_1|^2+|c_2|^2+2Re(\bar{c}_1c_2)\cos\theta_\pi^H
-\frac{3}{2}|c_1|^2\sin^2\theta_\pi^H\Big)\nonumber\\
&&\phantom{x}+\,\Big(2|c_1|^2\cos\theta_\pi^H+Re(\bar{c}_1c_2)\Big)
\sin2\Theta_\eta\sin\theta_\pi^H\cos\phi_\pi^H\nonumber\\
&&\phantom{x} -|c_1|^2\sin^2\Theta_\eta\sin^2\theta_\pi^H\cos2\phi_\pi^H\Big]\,,
\end{eqnarray}
where we omitted unessential kinematic factors.
The coefficients $c_1$ and $c_2$ in (\ref{75}) depend only on $\cos\theta_\pi^H$ and
$\omega_{\pi N}$, and are independent of $\Theta_\eta$ and $\phi^H_\pi$. In terms of the
functions $F^{D_{33}(\alpha)}$ (\ref{40a}) and (\ref{40b}) they read
\begin{eqnarray}\label{80}
c_1&=&\left(F^{D_{33}(\eta\Delta)}+F^{D_{33}(\pi N^*)}\right)q_\pi^*\,,\\
\label{80a} c_2&=&F^{D_{33}(\pi N^*)}X_\pi q_\eta\,.
\end{eqnarray}
The factor $X_\pi$ in eq.\,(\ref{80a}) comes from the Lorentz transformation to the $\pi
N$ rest frame $\vec{q}_\pi\to\vec{q}_\pi^{\,*}$ (see eq.\,(\ref{30})). The angular
distribution $W(\Omega_\pi^H)$ can easily be obtained from (\ref{75}) via appropriate
integration. Since $c_1$ and $c_2$ are independent of $\Theta_\eta$, the integration over
$\cos\Theta_\eta$ is trivial and we arrive at
\begin{equation}\label{85}
W(\theta_\pi^H)=A+B\cos\theta_\pi^H,
\end{equation}
with
\begin{eqnarray}
&&A=\frac{1}{N}\int\big(|c_1|^2+|c_2|^2\big)d\omega_{\pi N}\,,\nonumber\\
&&B=\frac{2}{N}\int Re(\bar{c}_1c_2)d\omega_{\pi N}\,,
\end{eqnarray}
where the normalization factor reads
\begin{eqnarray}\label{85N}
&&N=\int\big(|c_1|^2+|c_2|^2\nonumber\\
&&\phantom{xxxxx}+2Re(\bar{c}_1c_2)\cos\theta_\pi^H\big) d\omega_{\pi
N}d\cos\theta_\pi^H\,.
\end{eqnarray}
It is worth mentioning that the quantities $A$ and $B$ in eq.\,(\ref{85}) are still
functions of $\cos\theta_\pi^H$.

The distribution over the azimuthal angle $\phi_\pi^H$, obtained via integration of the
cross section (\ref{75}) over $\cos\Theta_\eta$, $\cos\theta_\pi^H$, and $\omega_{\pi N}$,
has the form
\begin{equation}\label{90}
W(\phi_\pi^H)=\frac{1}{2\pi}\left(1+\frac{1-a}{1+a}\ C\cos 2\phi_\pi^H\right)\,.
\end{equation}
The coefficient $C$ is independent of $\phi_\pi^H$ and reads
\begin{equation}\label{90C}
C=\frac{1}{N}\int |c_1|^2\sin^2\theta_\pi^H\,d\omega_{\pi N}\,d\cos\theta_\pi^H\,.
\end{equation}
The parameter $a$ is determined by the ratio of the $D_{33}$ helicity amplitudes as
\begin{equation}\label{90a}
a(W)=\left(\frac{A^{D_{33}}_{3/2}(W)}{A^{D_{33}}_{1/2}(W)}\right)^2\,.
\end{equation}

Several conclusions important for understanding the experimental data presented below can
be drawn. Firstly, as is found in ref.\,\cite{MAMI}, the distribution $W(\theta_\pi^H)$
is rather sensitive to the relative contribution of the $\pi N^*$ channel to the
$D_{33}(1700)$ decay. Quantitatively this sensitivity was discussed in \cite{MAMI} in
terms of the parameter
\begin{equation}\label{90r}
r=\frac{\Gamma^{(\pi N^*)}_{D_{33}\to\pi\eta N}}{\Gamma^{(\eta\Delta)}_{D_{33}\to\pi\eta
N}}\,,
\end{equation}
where $\Gamma^{(\alpha)}_{D_{33}\to\pi\eta N}$, $\alpha=\eta\Delta,\, \pi N^*$, stands for
the corresponding partial decay width.

From (\ref{80a}) and (\ref{85}) one sees that if the contribution of $\pi N^*$ vanishes,
$W(\theta_\pi^H)$ becomes isotropic ($B=0$). The addition of the nonzero $D_{33}\to\pi
N^*$ coupling results in an approximately linear dependence on $\cos\theta_\pi^H$ (there is
an additional dependence of the factors $F^{D_{33}(\pi N^*)}$ and $X_\pi$ in (\ref{80}) on
$\cos\theta_\pi^H$, which is, however, rather weak). The corresponding coefficient $B$ is
proportional to $Re(\bar{c}_1c_2)$ (\ref{85}) and is therefore rather sensitive even to a
small admixture of $\pi N^*$.

\begin{figure}
\begin{center}
\resizebox{0.3\textwidth}{!}{%
\includegraphics{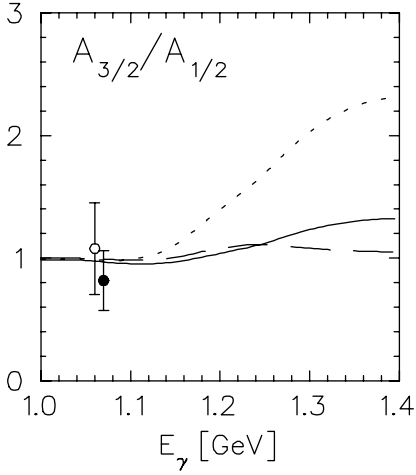}}
\caption{Ratio $A_{3/2}/A_{1/2}$ of the helicity amplitudes for $D_{33}(1700)$ as
a function of the lab photon energy. The solid, dashed, and dotted curves correspond to the
solutions I, II, and III from Table\,\ref{ta2}. The empty circle shows the ratio obtained
in ref.\,\cite{Arndt06}, whereas the filled circle is the value advocated by
PDG~\cite{PDG}.} \label{figa13}
\end{center}
\end{figure}

In the canonical frame one has
\begin{eqnarray}\label{100}
&&\frac{d\sigma}{d\omega_{\pi N}d\Omega_\pi^K d\cos\Theta_\eta}\sim \nonumber\\
&&\phantom{xxx}A^2_{1/2}\Big(|c_1|^2+|c_2|^2-2Re(\bar{c}_1c_2)\cos\theta_{\pi\eta}\Big)\nonumber\\
&&\phantom{xxx}+\left(A^2_{3/2}-A^2_{1/2}\right)\Big(|c_1|^2\sin^2\theta_\pi^K+
|c_2|^2\sin^2\Theta_\eta\nonumber
\\
&&\phantom{xxx}+ 2Re(\bar{c}_1c_2)\sin\theta_\pi^K\sin\Theta_\eta\cos\phi_\pi^K\Big)\,,
\end{eqnarray}
with $\theta_{\pi\eta}$ being the angle between $\vec{q}_\pi^{\,*}$ and $\vec{q}_\eta$.
Although the formula (\ref{100}) looks less cumbersome than eq.\,(\ref{75}), the
coefficients $c_1$ and $c_2$ are now functions of all four variables $\omega_{\pi N}$,
$\Omega_\pi^K$, and $\cos\Theta_\eta$. Therefore, a simple general analysis, like in the
case of the $H$ system, is not possible. The only conclusion we can draw is that since
$c_1$ and $c_2$ depend rather weakly on the angle $\phi_\pi^K$, the distribution
$W(\phi_\pi^K)$ can approximately be represented in the form
\begin{equation}\label{105}
W(\phi_\pi^K)\approx \frac{1}{2\pi}\left(1+\frac{1-a}{1+a}\
\tilde{C}\cos\phi_\pi^K\right)\,,
\end{equation}
where the coefficient $\tilde{C}$ is equal to
\begin{eqnarray}\label{110}
&&\tilde{C}=\frac{2}{N}\int Re(\bar{c}_1c_2)\\
&&\phantom{xxxxx???}\times\sin\Theta_\eta\sin\theta_\pi^K\, d\omega_{\pi
N}\,d\cos\Theta_\eta\, d\cos\theta_\pi^K\,.\nonumber
\end{eqnarray}
A look at eqs.\,(\ref{110}) and (\ref{90}) shows that, contrary to $W(\phi_\pi^H)$ whose
form depends only on the parameter $a$, in the $K$ frame the $\phi_\pi^K$-dependence is
solely due to interference between $\eta \Delta$ and $\pi N^*$ mechanisms. We therefore
expect that the shape of $W(\phi_\pi^K)$ is much more sensitive to the $\pi N^*$
contribution than the $\phi_\pi$-distribution in the helicity system. This assumption is
well confirmed by the analysis of ref.\,\cite{MAMI}.

\begin{table*}
\renewcommand{\arraystretch}{2.0}
\caption{Parameters of the isobar model fitted to the data in Fig.\,\ref{fig4}. The first
three rows give the parameter sets, to which we refer in the text as solutions I, II, and
III. The forth row lists the corresponding numbers given by the Particle Data
Group~\cite{PDG}. The sign of $\sqrt{\beta_{\eta\Delta}}A_{1/2}$ is the sign of the
product of $A_{1/2}$ and the $R\to\eta\Delta$ coupling constant $f_{R\Delta\eta}$ (see
eq.\,\protect(\ref{40a})). The relative sign between $R\to\eta\Delta$ and $R\to\pi N^*$
couplings is taken positive for all resonances. The partial cross sections $\sigma^R$ are
calculated according to the approximate formula \protect(\ref{250a}) at $W=M_R$. An
extremely large (small) value of the ratio $r$ (see, e.g., those for $P_{31}(1750)$ and
$F_{35}(1905)$) means that the corresponding partial decay width
$\Gamma^{(\eta\Delta)}_{\pi\eta N}\,(\Gamma^{(\pi N^*)}_{\pi\eta N})$ is close to zero.
$\chi^2/N=2.76$, 3.06, and 3.39 for the solutions I, II, and III, respectively.}
\label{ta2}
\begin{center}
\begin{tabular*}{17.4cm}
{@{\hspace{0.2cm}}c@{\hspace{0.3cm}}|@{\hspace{0.4cm}}c
@{\hspace{0.5cm}}c@{\hspace{0.5cm}}c@{\hspace{0.5cm}}c
@{\hspace{0.7cm}}c@{\hspace{0.7cm}}c@{\hspace{0.7cm}}c @{\hspace{0.7cm}}c}
\hline\hline\noalign{\smallskip} $J^{\pi}[L_{2T2J}(M_R)]$ & $M_R$ & $\Gamma^R_{tot}(M_R)$
& $\sqrt{\beta_{\eta\Delta}}A_{1/2}(M_R)$ & $\frac{A_{3/2}(M_R)}{A_{1/2}(M_R)}$ &
$r=\frac{\Gamma^{(\pi N^*)}_{\pi\eta
N}(M_R)}{\Gamma^{(\eta \Delta)}_{\pi\eta N}(M_R)}$ & $\sigma^R(M_R)$ \\
  & [MeV] & [MeV] & [10$^{-3}$GeV$^{-1/2}$] &  &  & [$\mu$b] \\
\noalign{\smallskip}\hline\noalign{\smallskip}
$\frac32^-[D_{33}(1700)]$  &  $1701\pm1$ & $375$ & $10.6\pm0.2$ & $0.95\pm0.01$ & $0.70\pm0.03$ & $0.186$ \\
                           &  $1720$ & $375$ & $15.1$ & $0.95$ & $0.67$ & $0.384$ \\
                           &  $1709$ & $375$ & $14.8$ & $1.00$ & $0.67$ & $0.370$ \\
                           &  $1700$ & $300$ &        & $0.82$ &        &         \\
$\frac32^+[P_{33}(1600)]$  &  $1657\pm5$ & $350$ & $-0.10\pm0.011$  & $0.39\pm0.06$ & $1.78\pm0.32$ & $6.70\cdot 10^{-6}$ \\
                           &  $1636$ & $350$ & $0.17$  & $1.20$ & $169$ & $7.02\cdot 10^{-3}$ \\
                           &  $1670$ & $350$ & $0.18$  & $3.71$ & $71.4$ & $1.90\cdot 10^{-2}$ \\
                           &  $1600$ & $350$ &          & $0.39$&    &              \\
$\frac12^+[P_{31}(1750)]$  &  $1832\pm1$ & $400$ & $8.35\pm0.9$  &        & $0.22\pm0.11$ & $3.78\cdot 10^{-2}$ \\
                           &  $1820$ & $400$ & $9.93$ &        & $2.6\cdot 10^{-2}$  & $4.53\cdot 10^{-2}$ \\
                           &  $1826$ & $400$ & $40.2$ &        & $1.0\cdot 10^{-3}$ & $0.720$ \\
                           &  $1750$ &       &        &        &                     &         \\
$\frac52^+[F_{35}(1905)]$  &  $1873\pm4$ & $330$ & $-25.5\pm0.6$ & $-0.70\pm0.03$ & $(9.1\pm7.5)\cdot 10^{-3}$ & $0.506$ \\
                           &  $1890$ & $330$ & $-10.8$ & $-1.64$ & $1.7\cdot 10^{-2}$ & $0.231$ \\
                           &  $1890$ & $330$ & $9.3\cdot 10^{-4}$ & $-1.94$ & $1.1\cdot 10^{6}$ & $2.38\cdot 10^{-3}$ \\
                           &  $1890$ & $330$ &           & $-1.73$ &    &              \\
$\frac32^+[P_{33}(1920)]$  &  $1894\pm3$ & $200$ & $11.87\pm0.42$ & $1.15\pm0.06$ & $0.14\pm0.09$ & $0.319$ \\
                           &  $1842$ & $200$ & $15.5$ & $1.49$ & $0.89$ & $1.29$ \\
                           &  $1848$ & $200$ & $4.42$ & $1.62$ & $6.4$ & $0.462$ \\
                           &  $1920$ & $200$ &         & $0.575$ &         &         \\
$\frac32^-[D_{33}(1940)]$  &  $1870\pm1$ & $450$ & $19.9\pm1.1$ & $1.65\pm0.02$ & $0.22\pm0.03$ & $0.695$ \\
                           &  $2069$ & $450$ & $-37.6$ & $1.77$ & $0.10$ & $2.26$ \\
                           &  $1937$ & $450$ & $-7.8\cdot 10^{-3}$ & $5.09$ & $1.1\cdot 10^{6}$ & $1.04$ \\
                           &  $1940$ &       &                      &         &                    &         \\
\noalign{\smallskip}\hline\hline
\end{tabular*}
\end{center}
\end{table*}

In Fig.\,\ref{fig4} we demonstrate our results for different angular distributions. As
noted above, we choose as independent kinematic variables those associated with the
$\eta+(\pi N)$ partition. Namely, the results are presented for the distributions
$W(\theta_\pi^{K(H)})$ and $W(\phi_\pi^{K(H)})$ over the pion angle in the $K$ and
$H$ system (see Fig.\,\ref{fig2}), as well as for the distribution $W(\Theta_\eta)$
over
the angle of the $\eta$ momentum in the overall c.m.\ frame. The solid, long-dashed, and
dotted curves present three different solutions, providing an acceptable description of
the data. The resulting parameter sets corresponding to each solution are collected in
Table\,\ref{ta2}. The partial cross sections in the last column in this Table are
calculated according to formula (\ref{250a}) at the energy $W=M_R$, and give
approximate contributions of the corresponding resonances to the total cross section.

It is worth mentioning that the fits collected in Table\,\ref{ta2} do not necessarily
give the lowest value of $\chi^2$. Rather, we present them as three sample results which
describe the same data via different dynamical contents. Namely, solution I (first row in
Table\,\ref{ta2}, solid line in Fig.\,\ref{fig4}) is mostly consistent with our
assumption about $D_{33}$ dominance over the whole energy region considered. Here other
partial waves are of relatively minor importance. Solution II (second row, long-dashed
line in Fig.\,\ref{fig4}) describes the data at $E_\gamma>1.2$ GeV via an appreciably large
contribution of the resonance $P_{33}(1920)$. Finally, solution III (third row, dotted
curve in Fig.\,\ref{fig4}) contains a small admixture of the $d$-wave in
$D_{33}(1700)\to\eta\Delta$ decay, which is neglected in the first two cases.

The following conclusions can be drawn:

(i) The general properties of the data in Fig.\,\ref{fig4} (the number of minima and the
convexity direction) are roughly described with only the $D_{33}$ wave. This fact,
considered as a strong evidence of the $D_{33}$ dominance was already discussed in
ref.\,\cite{MAMI}. In particular, one can see that at energies above $E_\gamma=1.2$ GeV
the experimentally observed shape of $W(\theta_\pi^H)$ and $W(\phi_\pi^H)$ is rather
close to the predicted form $a+b\cos\theta_\pi^H$ and $c+d\cos 2\phi_\pi^H$
(eqs.\,(\ref{85}) and (\ref{90})). In the canonical frame the distribution
$W(\phi_\pi^K)$ is also in good agreement with the analytic form $A+B\cos\phi_\pi^K$
provided by the single $D_{33}$ model (see eq.\,(\ref{105})). As is shown in
ref.\,\cite{MAMI}, other partial waves fail to describe all the angular distributions
simultaneously.

\begin{figure}
\begin{center}
\resizebox{0.48\textwidth}{!}{%
\includegraphics{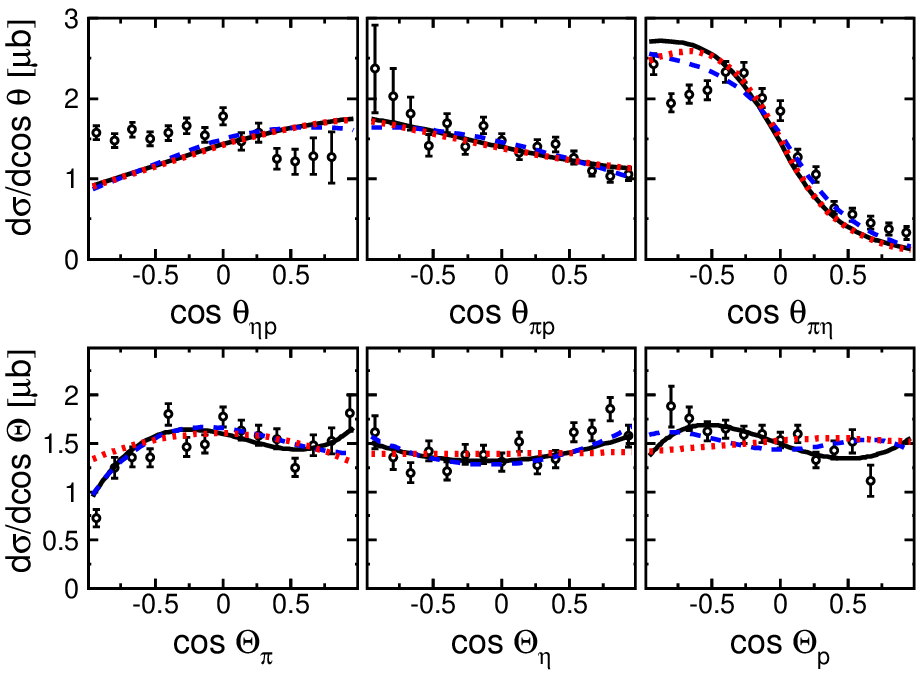}}
\caption{Our prediction of angular distributions for energy bin $1.7<W<1.9$\,GeV
compared with the CB-ELSA data of ref.\,\cite{Horn} (open circles).
In the first row, the angle $\theta_{ij}$ is the angle of the $i$th particle in the $ij$
c.m.\ frame with respect to the momentum of the third particle. In the second row,
$\Theta$ is the angle of the corresponding particle in the overall c.m.\ frame. The solid
(dashed) curve is obtained with the parameter set I (II) from Table
\protect\ref{ta2}. The dotted curve includes only the $D_{33}$ wave. The
theoretical results have arbitrary normalization.}
\label{figBonn}
\end{center}
\end{figure}

(ii) Since the coefficient $C$ in eq.\,(\ref{90}) is a constant, the $\phi_\pi^H$
dependence is totally determined by $\cos 2\phi_\pi^H$. Then, the convexity direction of
$W(\phi_\pi^H)$ (up or down) is governed by the sign of the difference $1-a$. Therefore,
the observed $\phi_\pi^H$-dependence allows one to make a conclusion about the value of the
parameter $a$. More specifically, according to the expression (\ref{90}), the
characteristic shape of $W(\phi_\pi^H)$ with the minima at $\phi_\pi^H=\pi/2$ and
$\phi_\pi^H=3\pi/2$, visible at $E_\gamma>1.2$ GeV, should point to the fact that at this
energy $a>1$ (eq.\,(\ref{90})). This result is also confirmed by $W(\theta_\pi^K)$ (first
row in Fig.\,\ref{fig4}) which, as follows from (\ref{105}) and (\ref{110}), would
otherwise be convex down in contrast to the data.

Our results for the ratio $A_{3/2}/A_{1/2}$, as a function of the photon energy, are
presented in Fig.\,\ref{figa13}. As may be seen, version III requires an unexpectedly
sharp energy dependence of this parameter, so that already at $E_\gamma=1.4$ GeV it
amounts up to 2.4, thus visibly exceeding the average PDG value $0.817\pm 0.242$. At the
same time, versions I and II allow one to suppress the undesirable increase of this
ratio at higher energies.

\begin{figure}
\begin{center}
\resizebox{0.49\textwidth}{!}{%
\includegraphics{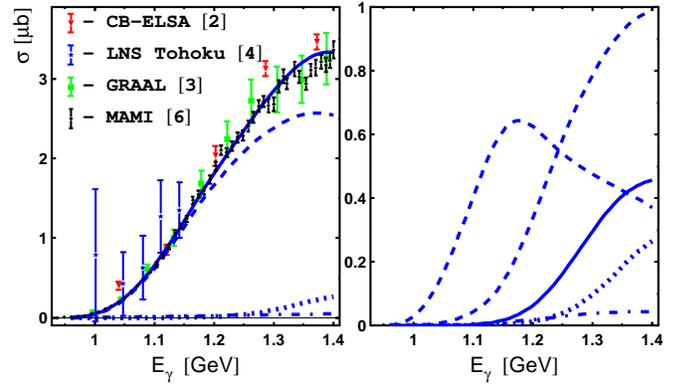}}
\caption{Total cross sections for $\gamma p\to\eta \pi^0 p$.
The theoretical results are obtained with
parameter set I in Table \protect\ref{ta2}. Left panel: full calculation (solid curve),
$D_{33}$ wave (dashed curve), $P_{33}$ wave (dotted curve), Born terms, shown
by the diagrams (a)-(f) in Fig.\,\protect\ref{fig1}, (dash-dotted curve). Right panel: the
contribution of $D_{33}(1700)$ and $D_{33}(1940)$ (dashed curves), $F_{35}(1905)$
(solid), $P_{33}(1920)$ (dotted), $P_{31}(1750)$ (dash-dotted).}
\label{fig5}
\end{center}
\end{figure}

(iii) According to our qualitative discussion above (see text after eq.\,(\ref{90r})),
the visible forward-backward asymmetry in $W(\theta^H_\pi)$ may be interpreted as a
signal of a nonzero $\pi N^*$ contribution. This conclusion is trivial in the sense that
the $\pi N^*$ channel always contributes to the reaction $\gamma p\to\pi^0\eta p$ via
$\eta p$ interaction in the final state. However as noted in the Introduction, direct
measurement of the $\eta p$ spectrum to study the role of this channel may fail,
primarily because of the overwhelming contribution of the $\eta\Delta$ decay mode. At the
same time, the role of $\pi N^*$ may be estimated rather precisely using the asymmetry
rate of $W(\theta^H_\pi)$. Indeed, as is shown in \cite{FOT}, here the different states
$J^\pi$ are added incoherently, so that the shape of $W(\theta^H_\pi)$ is always
symmetric if only the channel $\eta\Delta$ is taken into account. In this respect, the
observed asymmetry is a direct consequence of the interference between $\eta\Delta$ and
$\pi N^*$ mechanisms.


(iv) In the region $E_\gamma=1.2-1.4$\,GeV, the distribution $W(\Theta_\eta)$ over the
$\eta$ angle in the overall c.m.\ system demonstrates the characteristic shape
$a+b\cos^2\Theta$, vanishing at lower energies. This feature is explained in different
ways in our fits I-III. For example, the parameter sets I and II describe
$W(\Theta_\eta)$ via a relatively strong contribution of $F_{35}(1905)$ and
$P_{33}(1920)$ respectively . In the third version of the model, the increasing role of
the $\cos^2\Theta$ term is ascribed to the $d$-wave fraction in the $\eta\Delta$ decay of
$D_{33}(1700)$. Since the $D_{33}$ mass is pretty close to the $\pi\eta$ threshold, even
a small $d$-wave admixture in the $\eta\Delta$ mode may lead to a sizable contribution of
this wave at energies above $E_\gamma=1.2$ MeV. The dotted curve in Fig.\,\ref{fig4} is
obtained by assuming that the $d$-wave fraction comprises only 2$\%$ of the total
$D_{33}\to\pi\eta N$ width.

(v) In Fig.\,\ref{figBonn} we show our prediction of some angular distributions for three
different partitions, $\pi^0+(\eta p)$, $\eta+(\pi^0 p)$, and $p+(\pi^0\eta)$, together
with the data obtained in ref.\,\cite{Horn}. As one can see, our calculations are in
general agreement with the experimental results, perhaps except for the distribution over
$\cos\theta_{\eta p}$. At the same time, these observables demonstrate a rather weak
sensitivity to the model ingredients. Already the single $D_{33}$ model (short-dashed
line in Fig.\,\ref{figBonn}) correctly predicts the main features of all distributions.
Addition of other resonances does not visibly change the calculation. On the whole, all
three parameter sets in Table\,\ref{ta2} give equally good descriptions of the data.

(vi) As noted in the Introduction, the $D_{33}$ wave seems to also remain important at
higher energies. We base this conclusion on the fact that the single $D_{33}$ model works
quite well at $E_\gamma>1.3$ GeV (short-dashed line in Fig.\,\ref{fig4}). At the same
time, our prediction for the total cross section with only $D_{33}(1700)$ starts to
already underestimate the data at $E_\gamma=1.1$ GeV. This may be considered as an
indication of an additional source of the $D_{33}$ wave which should come into play at
higher energies. In ref.\,\cite{Horn} this effect is ascribed to $D_{33}(1940)$ which is
classified as a one-star resonance in the PDG listing~\cite{PDG}. In the present work, to
maintain the growth of the partial cross section $\sigma^{D_{33}}$ we also introduce this
resonance with free parameters.
However, one has to keep in mind that since $D_{33}(1940)$ is at the upper limit of our
energy region, the fitting results are not very sensitive to the exact contribution of
this resonance. The same must hold true for $P_{33}(1920)$.

(vii) Fig.\,\ref{fig5} demonstrates our calculation of the total cross section, where
contributions of individual terms entering the amplitude are plotted separately. The
results are obtained with the parameter set corresponding to the solution I in
Table\,\ref{ta2}. We prefer this version of our model since, as will be shown in the next
section, it also yields a reasonable description of the beam asymmetry, in contrast to the
solutions II and III. The reader should not be surprised by the quality of our
description in Fig.\,\ref{fig5} since, as noted in Sect.\,\ref{basis}, the total cross
section was effectively included into the fit (see the comment after eq.\,(\ref{55a})).

As predicted by solution I, the major fraction of the reaction yield is provided by
the coherent sum of the resonances $D_{33}(1700)$ and $D_{33}(1940)$ (short-dashed line
on the left panel in Fig.\,\ref{fig5}). The contribution of $D_{33}(1940)$ is comparable
to that of $D_{33}(1700)$. As pointed out above, the need for the former resonance in our
calculation is mainly dictated by the necessity to prevent the $D_{33}$ wave from dying
out when $D_{33}(1700)$ rapidly falls-off above $E_\gamma>1.2$\,GeV. Otherwise the model
leaves room for $P_{33}(1920)$ leading to solution II. As follows from
Table\,\ref{ta2}, in this case the contribution of $P_{33}(1920)$ amounts up to 40\,$\%$
of the total cross section at $E_\gamma=1.4$\,GeV.

It is quite remarkable that the role of $P_{33}(1600)$ in our analysis turns out to be
insignificant. The reason is that in the near-threshold region some fraction of the
$p$-wave comes from the background. The corresponding $p$-wave admixture turns out to be
sufficient to obtain a more or less accurate description of the forward-backward
asymmetry in the angular distribution $W(\Theta_\eta)$ of $\eta$ me\-sons in the overall
c.m. system. At the same time, at higher energies, where the role of $D_{33}$ strongly
increases, the background is unable to give the necessary $p$-wave fraction. As a result,
our fit requires inclusion of other resonances with positive parity, in particular
$P_{31}(1750)$, $F_{35}(1905)$, as well as $P_{33}(1920)$.

\subsection{Beam asymmetry}\label{BeamAsmtry}

\begin{figure}
\begin{center}
\resizebox{0.5\textwidth}{!}{%
\includegraphics{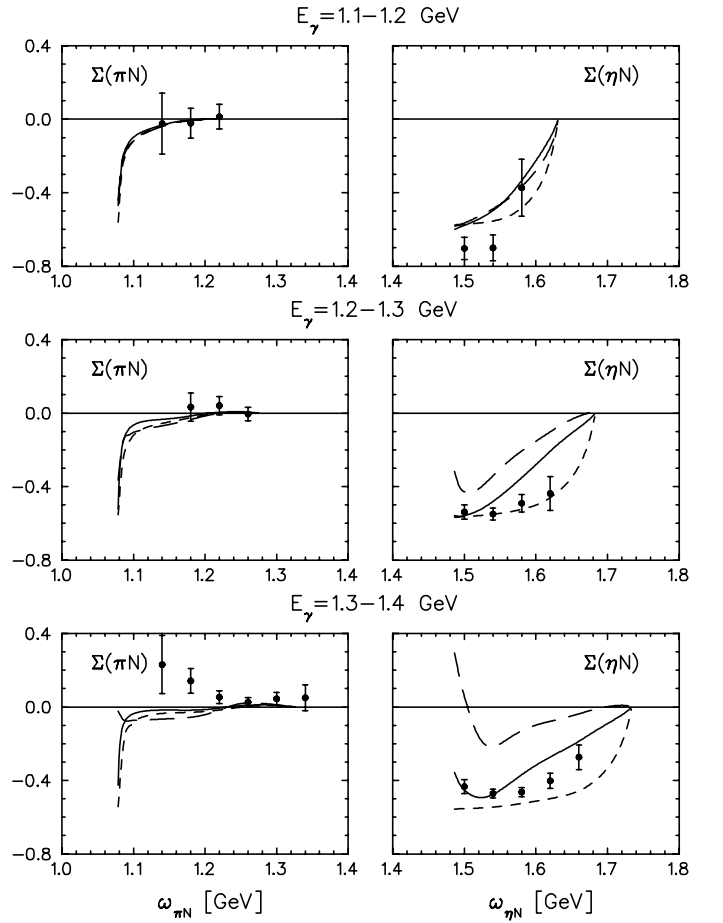}}
\caption{Beam asymmetry for $\gamma p\to\pi^0\eta p$ averaged over three intervals of the
beam energy. The data are from ref.\,\cite{Ajaka}. The solid (long-dashed) curve is our
prediction corresponding to the parameter set I (II) in Table \ref{ta2}. The short-dashed
curve includes only the $D_{33}$ wave.} \label{fig6}
\end{center}
\end{figure}

\begin{figure}
\begin{center}
\resizebox{0.48\textwidth}{!}{%
\includegraphics{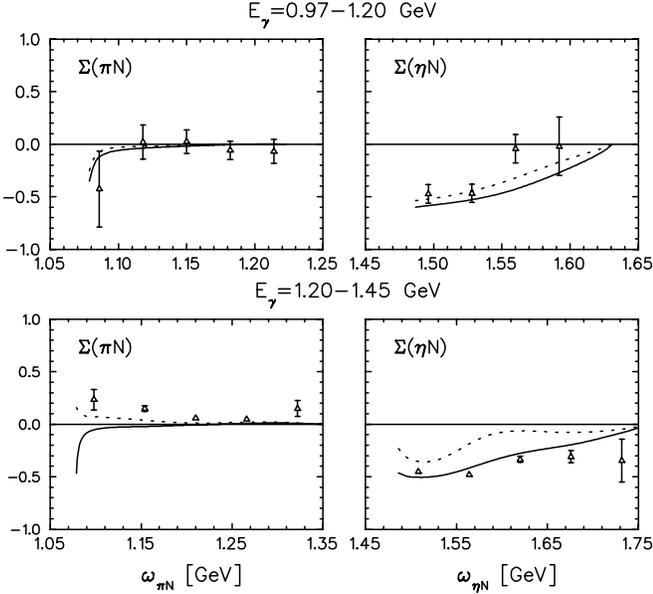}}
\caption{Same as in Fig.\,\protect\ref{fig6} but for other energy bins. The data are
taken from ref.\,\cite{Gutz}. The solid (dotted) curve corresponds to the solution I
(III) in Table \ref{ta2}.} \label{fig6a}
\end{center}
\end{figure}

The beam asymmetry $\Sigma$ for linearly polarized photons has been measured in
refs.\,\cite{Ajaka} and \cite{Gutz}. It was presented as a function of the invariant mass
of three two-particle subsystems $\pi^0p$, $\eta p$, and $\pi^0\eta$. In the region
$E_\gamma<1.5$ GeV this observable demonstrates several properties, which we would like
to discuss. Firstly, the data show (see Figs.\,\ref{fig6} and \ref{fig6a}) that in the
$\pi^0p$ case, when the reaction plane is fixed by the condition $\phi_\eta=0$ in the
overall c.m.\ system, the value of $\Sigma$ (hereafter denoted by $\Sigma(\pi N)$) is
comparable to zero in almost the whole region of $\omega_{\pi N}$, except for the lower
boundary of the spectrum. In the channel $\eta p$, where $\phi_\pi=0$, the beam asymmetry
$\Sigma(\eta N)$ is negative and its absolute value comprises about 0.5.

Both features may be qualitatively explained by the dominance of the $D_{33}$ partial
wave. Indeed, keeping only $D_{33}$ in the sum (\ref{20}), for the polarized cross
section one obtains (omitting unessential angular independent factors)
\begin{eqnarray}\label{20piNpol}
&&\Sigma\frac{d\sigma}{d\omega_{\pi
p}}=\frac{1}{2}\left(\frac{d\sigma_\perp}{d\omega_{\pi p}}
-\frac{d\sigma_\parallel}{d\omega_{\pi p}}\right) \nonumber\\
&&\phantom{xxxxxxx}\sim -\frac{A_{1/2}A_{3/2}}{\sqrt{3}}
\int\limits_0^\pi\sin\theta_\pi^H\,d\theta_\pi^H\nonumber\\
&&\phantom{xxxxxxx}\times\Big(|c_1|^2\frac{1}{2}\big(3\cos^2\theta_\pi^H-1)+|c_2|^2\nonumber\\
&&\phantom{xxxxxxx}+2\Re e(\overline{c}_1c_2)\cos\theta_\pi^H\Big)\,,
\end{eqnarray}
where the coefficients $c_i$ are determined in eqs.\,(\ref{80}) and (\ref{80a}). In
eq.\,(\ref{20piNpol}) $\sigma_\perp$ ($\sigma_\parallel$) is the cross-section with the
photon beam polarized perpendicular (parallel) to the reaction plane. The corresponding
unpolarized cross section reads
\begin{eqnarray}\label{20piN}
&&\frac{d\sigma}{d\omega_{\pi p}}=\frac{1}{2}\left(\frac{d\sigma_\perp}{d\omega_{\pi p}}
+\frac{d\sigma_\parallel}{d\omega_{\pi p}}\right) \nonumber\\
&&\phantom{xxxxx}\sim\frac{1}{2}\bigg(A_{1/2}^2+A_{3/2}^2\bigg)
\int\limits_0^\pi\sin\theta_\pi^H\,d\theta_\pi^H\nonumber\\
&&\phantom{xxxxx}\times\Big(|c_1|^2+|c_2|^2+2\Re
e(\overline{c}_1c_2)\cos\theta_\pi^H\Big)\,.
\end{eqnarray}
It is now straightforward to obtain an analytic expression for the beam asymmetry
dividing eq.\,(\ref{20piNpol}) by eq.\,(\ref{20piN}).

In the partition $\pi+(\eta N)$ the expressions for $\Sigma\frac{d\sigma}{d\omega_{\eta
N}}$ and for the unpolarized cross section $\frac{d\sigma}{d\omega_{\eta N}}$ are
formally identical to that given by (\ref{20piNpol}) and (\ref{20piN})
\begin{eqnarray}\label{20etaNpol}
&&\Sigma\frac{d\sigma}{d\omega_{\eta p}}
\sim -\frac{A_{1/2}A_{3/2}}{\sqrt{3}}
\int\limits_0^\pi\sin\theta_\eta^H\,d\theta_\eta^H\nonumber\\
&&\phantom{xxxx}\times\Big(|b_1|^2\frac{1}{2}\big(3\cos^2\theta_\eta^H-1)+|b_2|^2\nonumber\\
&&\phantom{xxxx}+2\Re e(\overline{b}_1b_2)\cos\theta_\eta^H\Big)\,,
\end{eqnarray}
\begin{eqnarray}\label{20etaN}
&&\frac{d\sigma}{d\omega_{\eta p}}
\sim\frac{1}{2}\bigg(A_{1/2}^2+A_{3/2}^2\bigg)
\int\limits_0^\pi\sin\theta_\eta^H\,d\theta_\eta^H\nonumber\\
&&\phantom{xxxx}\times\Big(|b_1|^2+|b_2|^2+2\Re
e(\overline{b}_1b_2)\cos\theta_\eta^H\Big)\,,
\end{eqnarray}
except that the integration variable is now $\theta^H_{\eta}$ and the coefficients $c_i$
are replaced by $b_i$, determined as
\begin{eqnarray}\label{bi}
b_1&=&-F^{D_{33}(\eta \Delta)}X_\pi q_\eta^*,\\
\label{bi1} b_2&=&\left(F^{D_{33}(\eta\Delta)}(1-X_\pi X_\eta)+F^{D_{33}(\pi
N^*)}\right)q_\pi.
\end{eqnarray}

With increasing $\omega_\alpha$, the value of $\Sigma(\pi N)$ rapidly goes to zero,
whereas $\Sigma(\eta N)$ demonstrates a rather smooth dependence. This qualitative
difference may be explained by the kinematic effect caused by the rather large mass
difference of the $\pi$ and $\eta$ mesons.

Indeed, because of the small mass of the pion, its momentum in the major part of the
reaction phase space is essentially higher than the $\eta$ momentum. As a consequence, in
the partition $\eta +(\pi N)$ we will have (see eqs.\,(\ref{80}) and (\ref{80a}))
\begin{equation}
\left|\frac{c_1}{c_2}\right|^{\,2}\sim\left(\frac{q_\pi^*}{q_\eta}\right)^2\gg 1\,.
\end{equation}
For example, taking for the parameter $r$ the value $r=0.7$, as predicted by the fit I,
we will have $|c_1/c_2|^2\approx 27$ at $W=1.7$ GeV, which further increases to 74 at
$W=1.9$ GeV.

\begin{figure}
\begin{center}
\resizebox{0.5\textwidth}{!}{%
\includegraphics{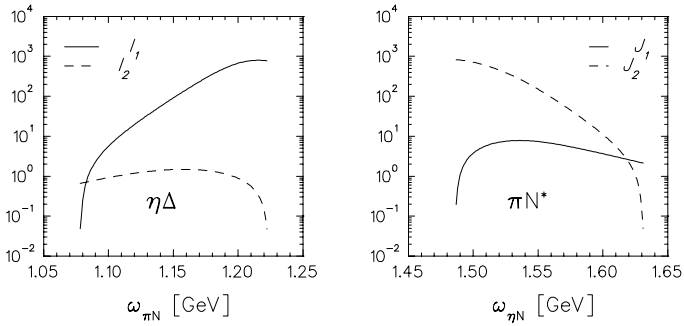}}
\caption{Integrals \protect(\ref{Ii}) and \protect(\ref{Ji}) calculated at photon energy
$E_\gamma$=1.2 GeV. The values are given in arbitrary units.} \label{fig7}
\end{center}
\end{figure}

Since in the cross sections (\ref{20piNpol}) and (\ref{20piN}) the coefficients $c_i$ appear
in the integrand, it is more appropriate to compare directly the
corresponding integrated values. As an example, in Fig.\,\ref{fig7} we show the integrals
\begin{eqnarray}\label{Ii}
I_i&=&\int\limits_0^\pi|c_i|^2\sin\theta_\pi^H\,d\theta_\pi^H\,,\quad i=1,2\,,\\
\label{Ji} J_i&=&\int\limits_0^\pi|b_i|^2\sin\theta_\pi^H\,d\theta_\pi^H\,,\quad i=1,2\,,
\end{eqnarray}
calculated at $E_\gamma=1.2$ GeV, as functions of the energy $\omega_\alpha$,
$\alpha=\eta\Delta,\,\pi N^*$. As noted above, in the region of very low $\omega_{\pi N}$
where $q_\pi^*\simeq 0$ we have $I_1\ll I_2$. However, with increasing $\omega _{\pi N}$
the pion momentum $q_\pi^*$ rapidly increases, so that the relation $I_1\gg I_2$ holds in
almost the whole region of $\omega_{\pi N}$. For instance, as may be seen from
Fig.\,\ref{fig7}, already at $\omega_{\pi N}=1.15$\,GeV we have $I_1/I_2\simeq 10^2$.

At the same time, in the polarized cross section $\Sigma \frac{d\sigma}{d\omega_{\pi N}}$
the coefficient $|c_1|^2$ enters with the weight $P_2(\cos\theta_\pi^H)=$
\\ $(3\cos^2\theta_\pi^H-1)/2$ (see eq.\,(\ref{20piNpol})), which strongly reduces its
contribution after integration over $\theta_\pi^H$. This last effect is due to the rather
smooth dependence of $|c_1|^2$ on $\cos\theta_\pi^H$. As a consequence, the value of
$\Sigma(\pi N)$ turns out to be roughly equal to
\begin{equation}
\Sigma(\pi N)\simeq \frac{I_2}{I_1+I_2}\simeq\frac{I_2}{I_1}\ll 1\,.
\end{equation}
In other words, in the region where the $D_{33}$ wave dominates, the nonzero value of the
asymmetry $\Sigma(\pi N)$ is due to the nontrivial contribution of the $\pi N^*$
configuration to the $D_{33}$ decay. For this reason, the rapid decrease of $\Sigma(\pi
N)$ at moderate $\omega_{\pi N}$ is due to the decrease of this contribution to the
integral in eq.\,(\ref{20piNpol}). The latter has a kinematic reason and, as we have just
explained, is eventually caused by the mass difference between $\pi$ and $\eta$ mesons.

In the second case, when the reaction plane is fixed by $\phi_\eta=0$ (partition $\eta
+(\pi N)$), the situation is just the opposite. Namely, (see
eqs.\,(\ref{bi})-(\ref{bi1}))
\begin{equation}
\left|\frac{b_1}{b_{\,2}}\right|^2\sim\left(\frac{q_\eta^*}{q_\pi} \right)^2\ll 1\,,
\end{equation}
so that
\begin{equation}
\left|\Sigma(\eta N)\right|\simeq \left|\frac{J_2}{J_1+J_2}\right|\simeq 1\,.
\end{equation}

In the same limiting case $\omega_{\eta N}\to 0$ ($b_1\to 0$), we get from
eqs.\,(\ref{20etaNpol}) and (\ref{20etaN}) a simple relation between the parameter $a$
(\ref{90a}) and the value of $\Sigma(\eta N)$
\begin{equation}\label{30Si}
\Sigma(\eta N)=-\frac{2}{\sqrt{3}}\,\frac{\sqrt{a}}{1+a}\,.
\end{equation}
Taking $a=1.2$ (the average of this parameter in the region $E_\gamma=1.0-1.4$ GeV) we
will have $\Sigma(\eta N)=-0.43$, which is consistent with the experimental result. Since
$\left|\Sigma(\eta N)\right|$ (\ref{30Si}) has maximum $\left|\Sigma(\eta
N)\right|=1/\sqrt{3}$ at $a=1$, it appears to be rather insensitive to $a$ as long as $a$
does not strongly differ from unity.

In a similar manner, at $\omega_{\pi N}\to 0$ we have $c_1\sim q_\pi^*\to 0$ and the
value of $\Sigma(\pi N)$ depends only on the parameter $a$\,(\ref{90a}). As a result, in
the single $D_{33}$ model (or generally in the single resonance model) one has the simple
relation
\begin{equation}\label{SiEq}
\Sigma(\pi N)\bigg|_{\,\omega_{\pi N}\to 0}=\Sigma(\eta N)\bigg|_{\,\omega_{\eta N}\to
0}\,.
\end{equation}

It is also worth noting that, according to the formula (\ref{20etaNpol}), the negative
value of the asymmetry $\Sigma(\eta N)$ is a signature of the fact that
$A^{D_{33}}_{1/2}$ and $A^{D_{33}}_{3/2}$ should have the same sign.

Our results for the photon asymmetry predicted by the three solutions discussed in the
previous section are presented in Figs.\,\ref{fig6} and \ref{fig6a}. As we can see, only
version I of our model where the $D_{33}$ wave dominates in the whole energy region
is able to reproduce the main features of the measured asymmetry. As would be expected,
the predicted behavior is close to that given by the single $D_{33}$ model, so that
inclusion of other resonances does not give rise to a crucial change. At the same time,
version II containing an appreciable contribution of $P_{33}(1920)$ (see the last column
in Table\,\ref{ta2}) yields a rather poor description of $\Sigma(\eta N)$ in the region
$1.3 \leq E_\gamma \leq 1.4$\,GeV. As one can see from Fig.\,\ref{fig6}, at low invariant
energies $\omega_{\eta N}$ this resonance causes an undesirable shift to positive
values, in contrast to the data demonstrating a rather smooth dependence on $\omega_{\eta
N}$.

\section{Conclusion}\label{Conclusion}

We performed a phenomenological analysis of the photon-induced $\pi^0\eta$ production at
energies within the region $1.0\le E_\gamma\le 1.4$\,GeV. The data contain detailed
information on angular distributions to determine, at least qualitatively, the
angular momentum content of the reaction amplitude.

As is repeatedly mentioned in the paper, among different mechanisms which might
contribute to the reaction, the excitation of the $D_{33}$ wave turns out to be most
evident in the data. According to our calculation, at the lab photon energy
$E_\gamma=1.1$ GeV and above the coherent sum of $D_{33}(1700)$ and $D_{33}(1940)$ can
account for the majority of the experimental results. We pay some attention to
a qualitative discussion of the way in which the main features of the angular distributions
determine some of the adjustable parameters of the model.

At the same time, deviations from the single $D_{33}$ model predictions point to
a nontrivial admixture of other resonance states. That these 'weaker' resonances should
have positive parity is confirmed by visible forward-backward asymmetry in the
$\cos\Theta_\eta$ distribution in the overall c.m.\ frame (fifth column in
Fig.\,\ref{fig4}).
Inclusion of $P_{33}(1600)$, $P_{31}(1750)$, and $F_{35}(1905)$, improves the quality of
our fit.

Furthermore, in order to maintain the rather monotone rise demonstrated by the measured
total cross section up to $E_\gamma=1.4$\,GeV we are forced to include the tails of the
resonance states with higher masses. Taking into account the results obtained in
ref.\,\cite{Horn}, the resonance $P_{33}(1920)$ as well as the one-star resonance
$D_{33}(1940)$ were included into the model. We allow their parameters to be varied,
although the energy region covered by the measurements does not permit an unambiguous
determination of their values.

On the whole, our results, at least those corresponding to solution I, are rather close
to the results obtained in ref.\,\cite{Horn}.
We also see that a
more rigorous study of the partial wave structure of $\pi^0\eta$ photoproduction requires
detailed polarization data. Otherwise, the theory admits a variation of the resonance
parameters over a relatively wide range. As an example, we give three versions of the fit
which explain the features of the observables via different mechanisms. In particular,
the characteristic shape of $W(\Theta_\eta)$ at higher energies, with a minimum at
$\cos\Theta_\eta=0$, may be described via different dynamical contents. In version III
this shape is reproduced if we allow for a small admixture of the $d$-wave mode in the
$\eta\Delta$ decay of the $D_{33}(1700)$. In the other two cases the same effect was
provided by the contribution of $P_{33}$ and/or $F_{35}$ states.

We also explored in considerable detail the photon asymmetry $\Sigma$, and found that its
main features may well be interpreted in terms of the single $D_{33}$ model. In this
model the nontrivial value of $\Sigma(\pi N)$ is a simple consequence of $\eta p$
interaction in the final state. Inclusion of other resonances leads to rather small
corrections. However, as pointed out in Sect.\,\ref{BeamAsmtry}, the behavior of
$\Sigma(\eta N)$ in the region of low $\eta N$ invariant energies is quite sensitive to
the contribution of the $P_{33}$ wave. In particular, our solution II containing a large
contribution of $P_{33}(1920)$ turns out to be inconsistent with the available data for
$\Sigma(\eta N)$.

Further study of the reaction $\gamma p\to\pi^0\eta p$ may be aimed at those observables,
in which the contributions of positive parity resonances may be identified in a model
independent way, wherever possible. With new data for polarization observables one may
expect considerable improvement of accuracy in the determination of the resonance parameters.
In
the past year some progress has been made towards that goal in refs.\,\cite{Gutz_Is} and
\cite{DorPlrz}.

This work was supported by the Deutsche Forschungsgemeinschaft (SFB 443), DFG-RFBR (Grant
No. 09-02-91330), the European Community-Research Infrastructure Activity under the FP6
``Structuring the European Research Area'' programme (Hadron Physics, contract number
RII3-CT-2004-506078), and the RF Presidential Grant (No MD-2772.2007.2). A.~Fix would
like to thank the Institut f\"ur Kernphysik of the Johannes Gutenberg-Universit\"at Mainz
for the kind hospitality.

\end{document}